Krzysztof Piasecki

Poznan University of Economics

## Behavioural present value

### Abstract

Impact of chosen behavioural factors on imprecision present value is discussed here. The formal model of behavioural present value is offered as a result of this discussion. Behavioural present value is described here by fuzzy set. These considerations were illustrated by means of extensive numerical case study. Finally there are shown that in proposed model the return rate is given, as a fuzzy probabilistic set.

### Research problem

Let us take into account any securities which are the object of trading on the established financial market. Our considerations in this article confine only to this financial instrument.

Typically, the analysis of properties of any securities of paper is kept, as analysis of properties of its return rate. Any return rate is an increasing function of the future value and a decreasing function of present value. The future value of the given financial instrument is presented, as a random variable[1]. Distribution of this random variable is the formal image of uncertainty risk of securities. In this work we limit our discussion to examine the order set by the return rate. We will compare the return rates set for the same security by the various financial market players. About the considered financial market we will assume that it is strongly effective. In this situation, all market players assume the same future value of chosen security. Then the increasing order of return rates is equivalent to the decreasing order of present value. In this situation, we decide to study the order set by present value. Thanks to that here we will omit here troublesome discussion on the detailed form of the distribution of return rate[2].

---

[1] In the case of a security which is free from the uncertainty risk, future value is given as a random variable with single-point distribution

[2] Can be found for example in: T. Winkler-Drews [29]. K. Echaust and E. Tomasik [9], E. Tomasik [23], [25], E. Tomasik and K. Piasecki [20], E. Tomasik and T. Wikler-Drews [24].

The accumulated knowledge of the financial market is the unique premise for determining a substantially justified price $C_0$ of considered security. In formal market analyses this price plays a role of declared equilibrium price. For this reason, the value $C_0$ is briefly called the equilibrium price. In our considerations equilibrium price plays a role of the synthetic image of market knowledge. We assumed that financial market is strongly effective. In this situation, all market players take the same value $C_0$ of the equilibrium price. At the same time, all those market players observe the same value $\check{C}$ market price. Knowledge of both of these values is sufficient for rational justification of investment decisions. For the case

$$\check{C} < C_0 \tag{1}$$

Rational rules clearly suggest the purchase of considered securities. Such purchase is possible only, when an offer for sale of these securities comes out. Here a question, what premises the investor selling such securities is guided by is natural. This puts the question about reasons by which securities seller is guided. Similarly, for the case

$$\check{C} > C_0 \tag{2}$$

Rational rules clearly suggest the sale of the considered securities. Such sale is possible only, when an offer for purchase of these securities comes out. This puts the question about reasons by which securities buyer is guided.

The reply to two above questions can be only one. On an effective financial market the balance between supply and demand is given under irrational reasons. It is obvious that these reasons could have behavioural character.

On the other hand, any present value of capital is defined, as such present value which is equivalent to given future value of capital. Mentioned equivalence relation is a subjective relationship, because it largely depends from investor's susceptibility to internal and external behavioural factors. It follows that due to the interaction of behavioural factors present value of security can be deviated from its observable market price. In fact, states of behavioural environment are defined imprecisely. For this reason, the deviation of the present value from market price is at the imprecision risk. Then the present value is imprecisely defined. The imprecise present value we will call the behavioural present value (in brief BPV).

The main purpose of this article will propose a formal model BPV. This model should serve to explain the mechanism of maintaining balance between demand and supply on the effective financial market

## 1. Ambiguity of the behavioural present value

The imprecision risk results from the imprecision of applied definitions and associated with it inaccuracy of observations of current event. This means that the imprecision risk is associated with information about the present event. Thus it differs from the uncertainty risk which is associated with information about future events. From the other side of many researchers of this subject (e.g. Klir, [14]) discuss two components of imprecision. They say that in the general case imprecision consists of ambiguity and indistinctness. The ambiguity of information is interpreted as a lack of clear recommendation one alternative between the various given alternatives. Indistinctness of information we interpret, as the lack of explicit distinguishing amongst the given information and its negation.

By its nature, only ambiguous information may be indistinct. Therefore in the first step we limit our discussion to the case of ambiguous description of present value. The most common and simplest model of ambiguous numerical information is interval number

The subjective evaluation of the present value is ambiguous. Each of considered evaluation alternatives is called a potential present value (abbreviated PPV). In this situation, ambiguous description of BPV is reduced to the determination of such interval, where each element is interpreted as a PPV.

Let us begin our deliberations on BPV from discussion the case of financial balance when the market price $\check{C}$ coincides with the equilibrium price $C_0$. Then we have

$$\check{C} = C_0 \quad . \tag{3}$$

This balance is momentary. This fact requires determine the PPV value as as a number close to market price. The assumed scope of PPV variability characterizes by a specific investor's susceptibility to the influence of internal and external behavioural factors. Then each of investors determines the following values:

- $C_{min}$ lower bond of PPV assumed under conditions of financial equilibrium;

- $C_{max}$ upper bond of PPV assumed under conditions of financial equilibrium.

In the case of the financial equilibrium the investor must take into account the possibility of quotation down trend and the possibility of rise in quotation. In this situation, the scope of PPV variability satisfies the condition

$$C_{min} < C_0 < C_{max}. \qquad (4)$$

Numerical $[C_{min}, C_{max}]$ is a BPV image for the case of the financial equilibrium.

Further discussion of BPV will guide now for the case when any market price $\check{C}$ is quoted. He is obvious, that BPV should be dependent on market price deviation from the equilibrium price

$$\Delta C = \check{C} - C_0 \qquad (5)$$

Then each of investors determines the following values:

- $\check{C}_{min}$ lower bound of PPV assumed for the case of market price $\check{C}$,
- $\check{C}_{max}$ upper bound of PPV assumed for the case of market price $\check{C}$.

Both of these values are dependent on the number $\alpha \in [0; 1]$ which determining the degree of investor's susceptibility to changes. The value of this degree tells us how the deviation $\Delta C$ strongly influences investor in his convictions change. This means that the value $1 - \alpha$ describes the degree of influence of the phenomenon of cognitive conservatism[3]. This phenomenon is taken into account in many behavioural models of the financial market. Discussion on this topic can be found for example in [1]. The degree of susceptibility is an individual investor's characteristic having behavioural base.

The investor determines the PPV lower scope, as the weighted average of lower bound $C_{min}$ assumed in terms of financial equilibrium and the value $C_{min} + \Delta C$ describing the same bound corrected by market price deviation from equilibrium one. In this equality the weight of the corrected lower bound is equal of the degree of the investor's susceptibility to changes. In determining the PPV

---

[3] Cognitive conservatism has been described on the psychology basis by W. Edwards [11]

lower boud investor must take into account the fact that this bound is always less or equal than the current market price. Thus we have here

$$\check{C}_{min} = min\{\alpha(C_{min} + \Delta C) + (1 - \alpha)C_{min}, \check{C}\} =$$
$$= min\{C_{min} + \alpha\Delta C, C_0 + \Delta C\}. \tag{6}$$

The investor determines the PPV upper bound, as the weighted average of upper range $C_{max}$ assumed in terms of financial equilibrium and the value $C_{max} + \Delta C$ describing the same bound corrected by market price deviation from equilibrium one. In this equality the weight of the corrected upper bound is equal of the degree of the investor's susceptibility to changes. In determining the PPV upper bound investor must take into account the fact that this bound is always greater or equal than the current market price. We have here

$$\check{C}_{max} = max\{\alpha(C_{max} + \Delta C) + (1 - \alpha)C_{max}, \check{C}\} =$$
$$= max\{C_{max} + \alpha\Delta C, C_0 + \Delta C\}. \tag{7}$$

Final estimates of scope of PPV variability is obtained in the form

$$\check{C}_{min} = \begin{cases} C_o + \Delta C & \Delta C \leq \frac{C_{min} - C_0}{1 - \alpha} \\ C_{min} + \alpha\Delta C & \Delta C > \frac{C_{min} - C_0}{1 - \alpha} \end{cases}, \tag{8}$$

$$\check{C}_{max} = \begin{cases} C_{max} + \alpha\Delta C & \Delta C < \frac{C_{max} - C_0}{1 - \alpha} \\ C_0 + \Delta C & \Delta C \geq \frac{C_{max} - C_0}{1 - \alpha} \end{cases}. \tag{9}$$

The scope of PPV variability assumed in terms of financial equilibrium and the degree of investor's susceptibility determine the impact of behavioral conditions on the PPV ambiguity. Individual investors differ in their susceptibility to environmental influences behavioral. They may therefore differ among themselves both the degree of susceptibility to change, and the scopes of PPV variability.

It is easy to above noted that, when in the case

$$\Delta C \leq \frac{C_{min} - C_0}{1 - \alpha} \tag{10}$$

of big surplus of equilibrium price over market one the BPV excludes the possibility of quotation downtrend. Also in case

$$\Delta C \geq \frac{C_{max} - C_0}{1 - \alpha} \tag{11}$$

of big surplus of market price over equilibrium one the BPV model excludes the possibility of rise in quotation. Conditions (10) or (11) describe the financial market in which significant deviation of market price from equilibrium price occurs. Then there are only rationales for investor's decision making. Scope of behavioural reasons impact is determined by following condition

$$\frac{C_{min} - C_0}{1 - \alpha} < \Delta C < \frac{C_{max} - C_0}{1 - \alpha} \tag{12}$$

fulfilled by market states.

Finally, for each investor we can designate a specific scope of PPV variability. This scope is dependent, *inter alia*, on the market price deviation $\Delta C$ from the equilibrium price. For this reason this scope is determined, as the function value $\mathbb{Z}(\Delta C)$. This function is given with the identity

$$\mathbb{Z}(\Delta C) = \begin{cases} [C_o + \Delta C; C_{max} + \alpha \cdot \Delta C] & for \ (10) \quad , \\ [C_{min} + \alpha \cdot \Delta C; C_{max} + \alpha \cdot \Delta C] & for \ (12) \quad , \\ [C_{min} + \alpha \cdot \Delta C; C_o + \Delta C] & for \ (11) \quad . \end{cases} \tag{13}$$

In this way we determine scope of present values expected under the influence of the current market situation.

## 2. Indistinctness of the behavioural present value

The interval image of BPV treats all acceptable PPV values in the same way. O the other side however we can suppose that the investor accepts PPV closer for market price more absolutely. It implies that individual PPV differs in their degrees of acceptance. This means that the interval model of BPV describes the complexity of the behavioural effects in insufficient way. This makes it necessary to build BPV model taking into account the variability of individual PPV importance. This leads directly to build an indistinct BPV model. The most frequent and simplest model of indistinct information is a fuzzy set[4]. In this situation any indistinct BPV may be described by its membership function assigning degree of acceptance to each PPV.

---

[4] The approach to the presentation of financial values as fuzzy sets derived from the I.J. Buckley [2] and M. L. Calzi [3]

We keep our further discussion for given value $\Delta C$ of market price deviation from equilibrium price. Then the scope $\mathbb{Z}(\Delta C) = \left[\check{C}_{min}; \; \check{C}_{max}\right]$ of PPV variability is determined explicitly. To unify further consideration, this interval will give up standardization. We use here following transformation

$$\beta = \begin{cases} \frac{p - \check{c}}{\check{c} - \check{c}_{min}} & p \in ]\check{c}_{min}; \; \check{C}] \\ \frac{p - \check{c}}{\check{c}_{max} - \check{c}} & p \in [\check{C}; \; \check{c}_{max} \; [ \end{cases}. \tag{14}$$

Let us notice, that according to relations (8) and (9), we transform here the variability scope $\left[\check{C}_{min}; \; \check{C}_{max}\right]$ into the standardized interval $\mathbb{I}(\Delta C)$ given by the identity

$$\mathbb{I}(\Delta C) = \begin{cases} [0; 1] & for \; (10) \quad , \\ [-1; 1] & for \; (12) \quad , \\ [-1; 0] & for \; (11) \quad . \end{cases} \tag{15}$$

The $|\beta|$ value determines the PPV relative distance from the equilibrium price. The degree $\gamma$ of PPV similarity to market price we define by the equality

$$\gamma = 1 - |\beta| \; . \tag{16}$$

The defined above degree of similarity $\gamma$ simultaneously determines the PPV relative distance from border of scope.

We determine the standardized model of imprecise BPV, as the fuzzy subset in standardized interval $\mathbb{I}(\Delta C)$. This subset is determined by its membership function $\nu(\cdot \, |\Delta C) : \mathbb{I}(\Delta C) \to [0; 1]$ which determines for each PPV its degree of acceptance. For this reason it is called distribution of acceptance. This distribution should be nondecreasing function of the degree of PPV similarity to market price. We will assume further that for the case of financial equilibrium (3) membership function $\nu(\cdot \, |0) : \mathbb{I}(0) \to [0; 1]$ is described by the identity

$$\nu(\cdot \, |0) = \nu_o(\cdot), \tag{17}$$

where $\nu_o : [-1; 1] \to [0; 1]$ is a given function additionally fulfilling the conditions:

–     $\nu_o(0) = 1 - \nu_o(-1) = 1 - \nu_o(1) = 1;$     (18)

–   $\nu_o$ is nondecreasing function on interval $[-1; 0]$;

- $v_o$ is nonincreasing function on interval $[0; 1]$.

Condition (18) announces, that in the case (3) the equilibrium price is fully accepted, as permissible PPV. Assumed monotonicity of the function $v_o$ means that with the increasing degree of PPV similarity to the market price its acceptance degree doesn't decrease. Distribution of acceptance $v(\cdot\,|0)$ will be the reference point for determining any distribution of acceptance in the case of disequilibrium (1) or disequilibrium (2). Therefore it will be called the reference distribution of acceptance.

The second reference point for determining any distribution of acceptance will be rational forecast $\Theta(\cdot\,|\Delta C): [-1; 1] \to \{0; 1\}$ changes in quotation. It is known, that:

- if the disequilibrium condition (1) is fulfilled, then rationale require us to expect rise in the quotation,
- if the disequilibrium condition (2) is fulfilled, then rationale require us to expect quotation downtrend,
- if the balance condition (3) is fulfilled, then rationale require us to expect quotation downtrend or a rise in the quotation.

Thus, the function describing the rational forecast is given by the identity

- for $\Delta C < 0$

  - $$\Theta(\beta|\Delta C) = \begin{cases} 0 & \beta < 0 \\ 1 & \beta \geq 0 \end{cases},$$ (19)

- for $\Delta C > 0$

$$\Theta(\beta|\Delta C) = \begin{cases} 1 & \beta \leq 0 \\ 0 & \beta > 0 \end{cases},$$ (20)

- for $\Delta C = 0$

$$\Theta(\beta|0) = 1 \quad .$$ (21)

The importance of rational forecasts increases with the increase in market price distance $|\Delta C|$ from the equilibrium price and with the increase in the PPV distance $\gamma$ from scope border. This means that the pair of the market price and the scope border is a reference point for assessing the significance of rational forecasts. In this situation the product $\gamma \cdot |\Delta C|$ assesses the distance from this reference point.

For any deviation $\Delta C$ investor assesses the degree of acceptance as a weighted average of rational forecast and reference distribution of acceptance. Discussion about stimulants of rational forecast importance justifies determination its weight, as its normalized distance from the reference point. Then the distribution of acceptance is described by the identity

$$\nu(\beta|\Delta C) = \frac{1}{1 + \gamma \cdot |\Delta C|} \cdot \nu_o(\beta) + \frac{\gamma|\Delta C|}{1 + \gamma \cdot |\Delta C|} \cdot \Theta(\beta|\Delta C) =$$

$$= \frac{1}{1 + (1-|\beta|) \cdot |\Delta C|} \cdot \nu_o(\beta) + \frac{(1-|\beta|) \cdot |\Delta C|}{1 + (1-|\beta|) \cdot |\Delta C|} \cdot \Theta(\beta|\Delta C). \tag{22}$$

The described above membership function describes the standardized model BPV. Using the inverse of the transformation (14) we appoint now the BPV model for general case. In general, this model is a fuzzy subset in the variability scope $\mathbb{Z}(\Delta C) = [\check{C}_{min}; \check{C}_{max}]$. Its membership function $\mu(\cdot|\Delta C): \mathbb{Z}(\Delta C) \to [0; 1]$ is defined as follows

$$\mu(p|\Delta C) = \begin{cases} \nu\left(\frac{p-\check{c}}{\check{c}-\check{c}_{min}}\Big|\Delta C\right) & p \in [\check{C}_{min}; \ \check{C}] \neq \{\check{C}\} \\ \nu\left(\frac{p-\check{c}}{\check{c}_{max}-\check{c}}\Big|\Delta C\right) & p \in [\check{C}; \check{C}_{max}] \neq \{\check{C}\} \end{cases}. \tag{23}$$

The main goal of building a formal model BPV is an explanation for maintaining of balance between supply and demand on efficient financial market. In order to solve this problem, we express explicitly the influence of the market price deviation $\Delta C$ on the membership function $\mu(\cdot|\Delta C)$. Taking into account together (8), (9), (14), (22) and (23) we obtain following results:

– if (10) is fulfilled then for $p \in [C_0 + \Delta C; C_{max} + \alpha \cdot \Delta C]$ we have

$$\mu(p|\Delta C) = \lambda(p|\Delta C), \tag{24}$$

– if (12) is fulfilled then for $p \in [C_{min} + \alpha \cdot \Delta C; C_{max} + \alpha \cdot \Delta C]$ and $\Delta C < 0$ we have

$$\mu(p|\Delta C) = \begin{cases} \kappa(p|\Delta C) & p < C_0 + \Delta C \\ \lambda(p|\Delta C) & p \geq C_0 + \Delta C \end{cases}, \tag{25}$$

- if (12) is fulfilled  then for  $p \in [C_{min} + \alpha \cdot \Delta C; C_{max} + \alpha \cdot \Delta C]$  and $\Delta C \geq 0$ we have

$$\mu(p|\Delta C) = \begin{cases} \varphi(p|\Delta C) & p < C_0 + \Delta C \\ \psi(p|\Delta C) & p \geq C_0 + \Delta C \end{cases}, \tag{26}$$

- if (11) is fulfilled  then for  $p \in [C_{min} + \alpha \Delta C; C_0 + \Delta C]$  we have

$$\mu(p|\Delta C) = \varphi(p|\Delta C) \,, \tag{27}$$

Used above functions $\kappa(p|\Delta C)$, $\lambda(p|\Delta C)$, $\varphi(p|\Delta C)$, $\psi(p|\Delta C)$ are described by following relationships

- for $p \in \left[ \check{C}_{min}; \; \check{C} \right]$

$$\kappa(p|\Delta C) = \frac{\check{c} - \check{c}_{min}}{\check{c} - \check{c}_{min} + (p - \check{c}_{min}) \Delta c} \cdot v_o \left( \frac{p - \check{c}}{\check{c} - \check{c}_{min}} \right) \,, \tag{28}$$

$$\varphi(p|\Delta C) = \frac{\check{c} - \check{c}_{min}}{\check{c} - \check{c}_{min} + (p - \check{c}_{min}) \Delta c} \cdot \left( v_o \left( \frac{p - \check{c}}{\check{c} - \check{c}_{min}} \right) + \frac{p - \check{c}_{min}}{\check{c} - \check{c}_{min}} \Delta C \right), \tag{29}$$

- for $p \in \left[ \check{C}; \check{C}_{max} \right]$

$$\lambda(p|\Delta C) = \frac{\check{c}_{max} - \check{c}}{\check{c}_{max} - \check{c} + (\check{c}_{max} - p) \Delta c} \cdot \left( v_o \left( \frac{p - \check{c}}{\check{c}_{max} - \check{c}} \right) + \frac{\check{c}_{max} - p}{\check{c}_{max} - \check{c}} \Delta C \right) \,, \tag{30}$$

$$\psi(p|\Delta C) = \frac{\check{c}_{max} - \check{c}}{\check{c}_{max} - \check{c} + (\check{c}_{max} - p) \Delta c} \cdot v_o \left( \frac{p - \check{c}}{\check{c}_{max} - \check{c}} \right). \tag{31}$$

Established in this article, the main goal of building a formal BPV model was an attempt to clarify the mechanism of maintaining balance between demand and supply on efficient financial market. In order to solve this problem, we present explicitly the impact of market price deviations $\Delta C$ on membership function $\mu(\cdot \,|\Delta C) \colon \mathbb{Z}(\Delta C) \to [0; 1]$. We have here

- if (10) is fulfilled then for $p \in [C_0 + \Delta C; C_{max} + \alpha \Delta C]$

$$\mu(p|\Delta C) = \lambda(p|\Delta C) \,, \tag{32}$$

- if (12) is fulfilled  and  $\Delta C < 0$ then

$$\mu(p|\Delta C) = \begin{cases} \kappa(p|\Delta C) & p \in [C_{min} + \alpha \Delta C; \; C_0 + \Delta C] \\ \lambda(p|\Delta C) & p \in [C_0 + \Delta C; C_{max} + \alpha \Delta C] \end{cases} \,, \tag{33}$$

- if (12) is fulfilled  and  $\Delta C \geq 0$ then

$$\mu(p|\Delta C) = \begin{cases} \varphi(p|\Delta C) & p \in [C_{min} + \alpha \cdot \Delta C; \ C_0 + \Delta C] \\ \psi(p|\Delta C) & p \in [C_0 + \Delta C; C_{max} + \alpha \Delta C] \end{cases}, \tag{34}$$

– if (11) is fulfilled then for $p \in [C_{min} + \alpha \Delta C; C_0 + \Delta C]$

$$\mu(p|\Delta C) = \varphi(p|\Delta C), \tag{35}$$

and

- for $p \in [C_{min} + \alpha \Delta C; \ C_0 + \Delta C]$

$$\kappa(p|\Delta C) = \frac{c_0 - c_{min} + (1-\alpha)\Delta C}{c_0 - c_{min} + (1-\alpha)\Delta C + (p - c_{min} - \alpha \Delta C) \cdot \Delta C} \cdot \nu_o \left( \frac{p - (c_0 + \Delta C)}{c_0 - c_{min} + (1-\alpha)\Delta C} \right), \tag{36}$$

$$\varphi(p|\Delta C) = \kappa(p|\Delta C) \cdot \left( 1 + \frac{p - c_{min} - \alpha \Delta C}{c_0 - c_{min} + (1-\alpha)\Delta C} \cdot \frac{\Delta C}{\nu_o \left( \frac{p - (c_0 + \Delta C)}{c_0 - c_{min} + (1-\alpha)\Delta C} \right)} \right), \tag{37}$$

- for $p \in [C_0 + \Delta C; C_{max} + \alpha \Delta C]$

$$\psi(p|\Delta C) = \frac{c_{max} - c_0 + (\alpha - 1)\Delta C}{c_{max} - c_0 + (\alpha - 1)\Delta C + (c_{max} + \alpha \Delta C - p) \cdot \Delta C} \cdot \nu_o \left( \frac{p - (c_0 + \Delta C)}{c_{max} - c_0 + (\alpha - 1)\Delta C} \right), \tag{38}$$

$$\lambda(p|\Delta C) = \psi(p|\Delta C) \cdot \left( 1 + \frac{p - c_{max} - \alpha \Delta C}{c_{max} - c_0 + (\alpha - 1)\Delta C} \cdot \frac{\Delta C}{\nu_o \left( \frac{p - (c_0 + \Delta C)}{c_{max} - c_0 + (\alpha - 1)\Delta C} \right)} \right), \tag{39}$$

This model is a fuzzy number according to J. Dubois and H. Prade [8]. Average value of this number may be interpreted as the average PPV. For each market price deviation $\Delta C$ we can assign a value $\xi(\Delta C)$ of the average PPV. This value is given by the identity

$$\xi(\Delta C) = \left( \int_{\mathbb{Z}(\Delta C)} \mu(x|\Delta C) dx \right)^{-1} \cdot \int_{\mathbb{Z}(\Delta C)} x \cdot \mu(x|\Delta C) dx. \tag{40}$$

The average PPV $\xi(\Delta C)$ determined for a particular investor can be interpreted as his average subjective evaluation of present value. An objective assessment of present value identified with the equilibrium price $C_0$ is here only one of reason which determines the subjective evaluation of present value. In this situation, from the investor's point-view the average PPV $\xi(\Delta C)$ is more reliable information than the equilibrium price $C_0$. This means that the investor's decisions are dependent on the relationship between the market price $\check{C}$ and the average PPV $\xi(\Delta C)$.

If the condition

$$\check{C} < \xi(\Delta C) \qquad (41)$$

is fulfilled then the investor recognizes that the financial market forces down the price of considered security. Therefore the investor expects quick increase in market price of the financial instrument. These expectations justify the submission of offer to buy considered security. The value of this demand depends on the investor's investment strategy and money stock on hand. If is met the conditions

$$\check{C} > \xi(\Delta C) \qquad (42)$$

is fulfilled then the investor recognizes that the financial market forces up the price of considered security. Therefore the investor expects quick decrease in the market price of this financial instrument. These expectations justify the submission of offer for sale considered security. The value of this supply is limited from above by value of securities on hand.

Let us notice here, that subjective condition (41) replaces the objective condition (1) and subjective condition (42) replaces the objective condition (2). It is obvious that on the effective financial market the conditions (1) and (2) could not be fulfilled at the same time.

The influence of behavioural environment on investor is described by the scope of PPV variability assumed by him under conditions of financial equilibrium, the degree of investor's susceptibility to changes and his reference distribution of acceptance. These characteristics are specific to each investor. From the other side all mentioned above factors are behavioural determinants of BPV. It means that on the effective financial market each investor determines his BPV in the specific way. In this situation, the variability of average PPV may be a specific characteristic of each investor. This means that on efficient financial market investors satisfying the condition (41) and investors satisfying the condition (42) can be found at the same time. In contrast to the conjunction of conditions (1) and (2), on efficient financial market conditions (33) and (34) can be simultaneously fulfilled. In this situation, demand caused by investors satisfying the condition (41) is balanced by the supply offered by the investors satisfying the condition (42). However we should remember that this observation applies only to investors present on the financial market at given moment. Reasons impelling the investor to enter given financial market are not included in proposed model.

If sales reduction or purchase reduction doesn't take place then observed market price $\check{C}$ is a market balance price in the sense given by microeconomic. Market balance price depends largely on investors' vulnerability to behavioural environment. The price $C_0$ is equilibrium price in financial sense. The financial equilibrium price describes the most stable price of securities. This means that on the financial market we can observe the financial equilibrium price $C_0$ and the market balance price $\check{C}$. These prices may be different. This conclusion fully explains the paradox of maintaining the market balance on the effective financial market.

Even superficially analysing the variability of acceptance distribution is easy to see that:

- condition (10) is a sufficient for (41) one;
- condition (11) is a sufficient for (42) one.

In this situation only a case (12) requires a detailed analysis.

### 3. Behavioural present value – a case of study

Complex form of the BPV model leads us to study its properties through computational experiments. The first step toward this type of numerical research is the following case study. The main goal of this case study will be a demonstration event in which offers to buy and offers for sale appear at the same time.

We consider a financial instrument characterized by equilibrium price $C_0 = 100$. Investor A and investor B are interested in participating in these securities trade.

Investor A's susceptibility to the influence of internal and external behavioural factors is described by means of:

- $C_{min}^A = 95$ lower bound of PPV assumed under conditions of financial equilibrium;
- $C_{max}^A = 110$ upper bound of PPV assumed under conditions of financial equilibrium.

Investor B susceptibility to the influence of internal and external behavioural factors is described by means of with the value:

- $C_{min}^B = 90$ lower bound of PPV assumed under conditions of financial equilibrium;
- $C_{max}^B = 105$ upper bound of PPV assumed under conditions of financial equilibrium.

Comparison of both of these scopes shows that, for the case of financial equilibrium (3), market expectations of the investor A is more optimistic than market expectations of the investor B.

Degree of the investor A susceptibility to changes is equal to $\alpha_A = 0{,}2$. The analogous degree of the investor B susceptibility is equal to $\alpha_B = 0{,}8$. It is evident here is that market reaction of investor B is quicker than the market reaction of investor B.

In this way, we gathered all information necessary to determine ambiguous BPV. It is easy to see that each investor has one advantage and one disadvantage. The advantages are more optimistic market expectation of investor A and faster market reaction of investor B. Disadvantages are the more pessimistic market expectations of investor B and slower market reaction of investor A.

Using the equality (13) we appoint the PPV variability scope for the investor A. We have here

$$\mathbb{Z}_A(\Delta C) = \begin{cases} [100 + \Delta C; 110 + 0{,}2 \cdot \Delta C] & \Delta C \leq -6{,}25 \ , \\ [95 + 0{,}2 \cdot \Delta C; 110 + 0{,}2 \cdot \Delta C] & -6{,}25 < \Delta C < 12{,}5, \\ [95 + 0{,}2 \cdot \Delta C; 100 + \Delta C] & \Delta C \geq 12{,}5. \end{cases} \quad (43)$$

For the investor B the PPV variability scope takes the form

$$\mathbb{Z}_B(\Delta C) = \begin{cases} [100 + \Delta C; 105 + 0{,}8 \cdot \Delta C] & \Delta C \leq -50 \ , \\ [90 + 0{,}8 \cdot \Delta C; 105 + 0{,}8 \cdot \Delta C] & -50 < \Delta C < 25, \\ [90 + 0{,}8 \cdot \Delta C; 100 + \Delta C] & \Delta C \geq 25. \end{cases} \quad (44)$$

We note here that if market price exceeds the level $\check{C}_A = 112{,}5$ then more optimistic investor A rules out further increases in quotation. For the same financial instrument the investor B rules out the possibility of the further increases in the quotation only in the casa when market price exceeds the level $\check{C}_B = 125$. The effect is due to the faster market reaction of the investor B.

Scopes of the PPV variability and degrees of the susceptibility to market changes are sufficient information for the designation components of BPV model described by equations (36), (37), (38) and (39).

For the investor A we obtain here following relationships:

- if $p \in [95 + 0{,}2 \cdot \Delta C; \; 100 + \Delta C]$ then

$$\kappa_A(p|\Delta C) = \frac{5+0{,}8\cdot\Delta C}{5+(p-94{,}2)\cdot\Delta C-0{,}2\cdot(\Delta C)^2} \, v_A\left(\frac{p-(100+\Delta C)}{5+0{,}8\cdot\Delta C}\right), \tag{45}$$

$$\varphi_A(p|\Delta C) = \frac{5+0{,}8\cdot\Delta C}{5+(p-94{,}2)\cdot\Delta C-0{,}2\cdot(\Delta C)^2}\left(v_A\left(\frac{p-(100+\Delta C)}{5+0{,}8\cdot\Delta C}\right) + \frac{p-95-0{,}2\cdot\Delta C}{5+0{,}8\cdot\Delta C}\cdot\Delta C\right) \;, \tag{46}$$

- if $p \in [100 + \Delta C; 110 + 0{,}2 \cdot \Delta C]$ then

$$\psi_A(p|\Delta C) = \frac{10-0{,}8\cdot\Delta C}{10+(109{,}2-p)\cdot\Delta C+0{,}2\cdot(\Delta C)^2} \, v_A\left(\frac{p-(100+\Delta C)}{10-0{,}8\cdot\Delta C}\right), \tag{47}$$

$$\lambda_A(p|\Delta C) = \frac{10-0{,}8\cdot\Delta C}{10+(109{,}2-p)\cdot\Delta C+0{,}2\cdot(\Delta C)^2}\left(v_A\left(\frac{p-(100+\Delta C)}{10-0{,}8\cdot\Delta C}\right) + \frac{p-110-0{,}2\cdot\Delta C}{10-0{,}8\cdot\Delta C}\cdot\Delta C\right). \tag{48}$$

where $v_A: [-1; 1] \longrightarrow [0; 1]$ is balanced distribution of acceptance determined by investor A.

For the investor B we obtain here following relationships:

- if $p \in [90 + 0{,}8 \cdot \Delta C; \; 100 + \Delta C]$ then

$$\kappa_B(p|\Delta C) = \frac{10+0{,}2\cdot\Delta C}{10+(p-89{,}8)\cdot\Delta C-0{,}8\cdot(\Delta C)^2} \cdot v_B\left(\frac{p-(100+\Delta C)}{10+0{,}2\cdot\Delta C}\right), \tag{49}$$

$$\varphi_B(p|\Delta C) = \frac{10+0{,}2\cdot\Delta C}{10+(p-89{,}8)\cdot\Delta C-0{,}8\cdot(\Delta C)^2}\left(v_B\left(\frac{p-(100+\Delta C)}{10+0{,}2\cdot\Delta C}\right) + \frac{p-90-0{,}8\cdot\Delta C}{10+0{,}2\cdot\Delta C}\cdot\Delta C\right), \tag{50}$$

- if $p \in [100 + \Delta C; 105 + 0{,}8 \cdot \Delta C]$ then

$$\psi_B(p|\Delta C) = \frac{5-0{,}2\cdot\Delta C}{5+(104{,}8-p)\cdot\Delta C+0{,}8\cdot(\Delta C)^2} \cdot v_B\left(\frac{p-(100+\Delta C)}{5-0{,}2\cdot\Delta C}\right), \tag{51}$$

$$\lambda_B(p|\Delta C) = \frac{5-0{,}2\cdot\Delta C}{5+(104{,}8-p)\cdot\Delta C+0{,}8\cdot(\Delta C)^2}\left(v_B\left(\frac{p-(100+\Delta C)}{5-0{,}2\cdot\Delta C}\right) + \frac{p-105-0{,}8\Delta C}{5-0{,}2\cdot\Delta C}\Delta C\right), \tag{52}$$

where $v_B: [-1; 1] \longrightarrow [0; 1]$ is balanced distribution of acceptance determined by investor B.

For determining the balanced distribution of acceptance suggestion the suggestions, we apply suggestions given by Fang Yong, Kin Keung Lai and Wang Shouyang [11]. These authors recommend the use triangular or trapezoidal fuzzy numbers in analysis of securities. According to this suggestion we will determine balanced distribution of acceptance as triangular fuzzy numbers. It means that for both investors reference distribution of acceptance are identical and given by the identity

$$\nu_A(\beta) = \nu_B(\beta) = 1 - |\beta|. \tag{53}$$

The condition (53) is equivalent for assumption, that under the conditions of financial equilibrium (3) the acceptance degree of PPV is equal to its degree of similarity to the market price.

Now for the investor A we have here

- if $p \in [95 + 0,2 \cdot \Delta C; \ 100 + \Delta C]$ then

$$\kappa_A(p|\Delta C) = \frac{p - 95 - 0,2 \cdot \Delta C}{5 + (p - 94,2) \cdot \Delta C - 0,2 \cdot (\Delta C)^2}, \tag{54}$$

$$\varphi_A(p|\Delta C) = \frac{(p - 95 - 0,2 \cdot \Delta C) \cdot (1 + \Delta C)}{5 + (p - 94,2) \cdot \Delta C - 0,2 \cdot (\Delta C)^2} \ , \tag{55}$$

- if $p \in [100 + \Delta C; 110 + 0,2 \cdot \Delta C]$ then

$$\psi_A(p|\Delta C) = \frac{110 + 0,2 \cdot \Delta C - p}{10 + (109,2 - p) \cdot \Delta C + 0,2 \cdot (\Delta C)^2}, \tag{56}$$

$$\lambda_A(p|\Delta C) = \frac{(110 + 0,2 \cdot \Delta C - p) \cdot (1 - \Delta C)}{10 + (109,2 - p) \cdot \Delta C + 0,2 \cdot (\Delta C)^2}. \tag{57}$$

For the B investor we have here

- if $p \in [90 + 0,8 \cdot \Delta C; \ 100 + \Delta C]$ then

$$\kappa_B(p|\Delta C) = \frac{p - 90 - 0,8 \cdot \Delta C}{10 + (p - 89,8) \cdot \Delta C - 0,8 \cdot (\Delta C)^2}, \tag{58}$$

$$\varphi_B(p|\Delta C) = \frac{(p - 90 - 0,8 \cdot \Delta C) \cdot (1 + \Delta C)}{10 + (p - 89,8) \cdot \Delta C - 0,8 \cdot (\Delta C)^2} \ , \tag{59}$$

- if $p \in [100 + \Delta C; 105 + 0,8 \cdot \Delta C]$ then

$$\psi_B(p|\Delta C) = \frac{105 + 0,2 \cdot \Delta C - p}{5 + (104,8 - p) \cdot \Delta C + 0,8 \cdot (\Delta C)^2} \ , \tag{60}$$

$$\lambda_B(p|\Delta C) = \frac{(105 + 0,2 \cdot \Delta C - p) \cdot (1 - \Delta C)}{5 + (104,8 - p) \cdot \Delta C + 0,8 \cdot (\Delta C)^2} \ . \tag{61}$$

In the next step, for each market price deviation $\Delta C$ satisfies the condition (12), we determine the membership functions BPV.

In case of the investor A we have here:

–    if $-6{,}25 < \Delta C < 0$ then for $p \in [95 + 0{,}2 \cdot \Delta C; 110 + 0{,}2 \cdot \Delta C]$

$$\mu_A(p|\Delta C) = \begin{cases} \frac{p - 95 - 0{,}2 \cdot \Delta C}{5 + (p - 94{,}2) \cdot \Delta C - 0{,}2 \cdot (\Delta C)^2} & p < 100 + \Delta C \\ \frac{(110 + 0{,}2 \cdot \Delta C - p) \cdot (1 - \Delta C)}{10 + (109{,}2 - p) \cdot \Delta C + 0{,}2 \cdot (\Delta C)^2} & p \geq 100 + \Delta C \end{cases}, \quad (62)$$

–   if $0 \leq \Delta C < 12{,}5$ then for $p \in [95 + 0{,}2 \cdot \Delta C; 110 + 0{,}2 \cdot \Delta C]$

$$\mu_A(p|\Delta C) = \begin{cases} \frac{(p - 90 - 0{,}8 \cdot \Delta C) \cdot (1 + \Delta C)}{10 + (p - 89{,}8) \cdot \Delta C - 0{,}8 \cdot (\Delta C)^2} & p < 100 + \Delta C \\ \frac{110 + 0{,}2 \cdot \Delta C - p}{10 + (109{,}2 - p) \cdot \Delta C + 0{,}2 \cdot (\Delta C)^2} & p \geq 100 + \Delta C \end{cases}, \quad (63)$$

For the B investor we obtain:

–   if $-50 < \Delta C < 0$ then for $p \in [90 + 0{,}8 \cdot \Delta C; 105 + 0{,}8 \cdot \Delta C]$

$$\mu_B(p|\Delta C) = \begin{cases} \frac{p - 90 - 0{,}8 \cdot \Delta C}{10 + (p - 89{,}8) \cdot \Delta C - 0{,}8 \cdot (\Delta C)^2} & p < 100 + \Delta C \\ \frac{(105 + 0{,}2 \cdot \Delta C - p) \cdot (1 - \Delta C)}{5 + (104{,}8 - p) \cdot \Delta C + 0{,}8 \cdot (\Delta C)^2} & p \geq 100 + \Delta C \end{cases}, \quad (64)$$

–   if $0 \leq \Delta C < 25$ then for $p \in [90 + 0{,}8 \cdot \Delta C; 105 + 0{,}8 \cdot \Delta C]$

$$\mu_B(p|\Delta C) = \begin{cases} \frac{(p - 90 - 0{,}8 \cdot \Delta C) \cdot (1 + \Delta C)}{10 + (p - 89{,}8) \cdot \Delta C - 0{,}8 \cdot (\Delta C)^2} & p < 100 + \Delta C \\ \frac{105 + 0{,}2 \cdot \Delta C - p}{5 + (104{,}8 - p) \cdot \Delta C + 0{,}8 \cdot (\Delta C)^2} & p \geq 100 + \Delta C \end{cases}. \quad (65)$$

How easy to note, investors A and B differ in determined by (12) scopes of behavioural premises impact.

For each of investors we appoint[5] values $\xi(\Delta C)$ of average PPV. Next applying the interval bisection method we solve the inequalities (33) or (34). After all we obtain two conclusions:

–   the investor A fulfils the condition (41) iff $\Delta C < 5{,}24$,
–   the investor B fulfils the condition (42) iff $\Delta C > -19{,}12$..

---

[5] There was used program Mathematica v. 8.0.0.0 license number L4719-1731

It means that if market price $\check{C} \in ]80{,}88; 105{,}24[$ then demand from investor A may be offset by the supply offered by the investor B.

In this way it stayed demonstrated possibility of using the proposed model to the description of the phenomenon of achieving the market balance in conditions of the financial imbalance.

Thus was demonstrated the possibility of applying proposed model to description the phenomenon of maintaining market balance under conditions of financial disequilibrium.

## 4. Impact of behavioural present value on return rate

Let us assume that time horizon $t > 0$ of investment is fixed. Then considered financial instrument is determined by two values:

– anticipated future value $V_t \in \mathbb{R}^+$ ,

– assessed present value $V_0 \in \mathbb{R}^+$.

The basic characteristics of benefits by ownership this instrument is a return rate $r_t$ given by the identity

$$r_t = r(V_0, V_t). \qquad (66)$$

In the special case we have here:

– simple return rate

$$r_t = \frac{V_t - V_0}{V_0} = \frac{V_t}{V_0} - 1. \qquad (67)$$

– logarithmic return rate

$$r_t = ln\frac{V_t}{V_0}. \qquad (68)$$

In the general case, the function: $r: \mathbb{R}^+ \times \mathbb{R}^+ \to \mathbb{R}$ is a decreasing function of the present value. Thanks to that for any future value $V_t$ we can determine inverse function $r^{-1}(\cdot, V_t): \mathbb{R} \to \mathbb{R}^+$.

In the classic approach to the problem of return rate determining the present value of investment is identified with the observed market price $\check{C}$ price what we write

$$V_0 = \check{C} . \qquad (69)$$

The future value of investment $V_t$ is at risk of uncertainty about future events. Formal model of this uncertainty is presentation future value, as a random variable $\tilde{V}_t: \Omega = \{\omega\} \longrightarrow \mathbb{R}^+$. The set $\Omega$ is a set of basic future events on the financial market. If return rate is determined by condition (69), then it is a random variable at uncertainty risk. This random variable is determined by the identity

$$\tilde{r}_t(\omega) = r\left(\check{C}, \tilde{V}_t(\omega)\right).$$

(70)

In practice of the financial markets analysis, the uncertainty risk is described by means of probability distribution of return rate. At the moment we have an extensive compendium of knowledge on this subject. Examples of this knowledge may be cited above papers [9], [20], [23], [24], [25] and [29]. Let us assume that probability distribution of return rate satisfying the condition (69) is given by the distribution function $F_r: \mathbb{R} \longrightarrow [0; 1]$. Then the probability distribution of future value is given by the distribution function $F_V: \mathbb{R}^+ \longrightarrow [0; 1]$ determined as follows

$$F_V(x) = F_r\left(r\left(\check{C}, \tilde{V}_t(\omega)\right)\right).$$

(71)

Cumulative distribution function $F_V: \mathbb{R} \longrightarrow [0; 1]$ describes the probability distribution of future value, which is value assessed ex post on the basis of objective measurement only. It means that the cumulative distribution function of future value is independent of the way of determining the present value.

As shown earlier, the present value may be subjected to imprecision risk. Let us assume that imprecise estimation of the present value is given as BPV described by its membership function $\mu_{BPV}: \mathbb{R}^+ \to [0; 1]$. Then the rate of return is subjected to uncertainty risk of future value and imprecision risk of present value. According to the Zadeh extension principle, for each fixed elementary state $\omega \epsilon \Omega$ of financial market, membership function $\rho(\cdot, \omega): \mathbb{R} \to [0; 1]$ of return rate is determined by the identity

$$\rho(r, \omega) = max\left\{\mu(y): y \epsilon \mathbb{R}^+, r = r\left(y, \tilde{V}_t(\omega)\right)\right\} = \mu\left(r_t^{-1}\left(r, \tilde{V}_t(\omega)\right)\right).$$

(72)

It means that considered return rate is represented by fuzzy probabilistic set. In special cases we have here:

– for the simple return rate

$$\rho(r,\omega) = \mu_{BPV}\left((1+r)^{-1} \cdot \tilde{V}_t(\omega)\right) \qquad (73)$$

– for the logarithmic return rate

$$\rho(r,\omega) = \mu_{BPV}\left(e^{-r} \cdot \tilde{V}_t(\omega)\right) \qquad (74)$$

It means that considered return rate is represented by fuzzy probabilistic set. The notion of probabilistic fuzzy set was suggested and studied by K. Hiroto [12]. For this reason, these sets are also called Hiroto's sets.

Despite this modernization, in the proposed model it is possible to apply the entire rich empirical knowledge collected about probability distribution of return rate. This is a highly advantageous feature of the proposed model, since it brings the possibility of real applications.

**Summary**

The research domain of behavioural finance is the paradoxes and anomalies in financial markets, which are difficult to explain based on neoclassical economic theory. Behavioural analysis of financial markets points to the psychological aspect of investment, as the reason for this state of affairs. At present extensive bibliography is already devoted to the results of these studies. On the Polish publishing market to the trend that we can include monographs M. J. Pring [21], T. Tyszka [28], T. Zaleskiewicz [30], P. Zielonka [31], A. Szyszka [22] M. Czerwonka and B. Gorlewski [5].

The consequence of this intensive research is aiming to such formal models which explain behavioural mechanisms of the financial market. Here we can distinguish a few approaches to this topic.

The most typical behavioural finance model is a formal prospect theory proposed by D. Kahneman and Tversky [26], [27]. In theory, a subjective transformation of objective probability is distinguished as a behavioural basis for investment decisions

N. Barberis, A. Shleifer and R. Visny [1] develop the prospect theory. They additionally point out imprecise estimation of present value, as a result of the subjective approach to security valuation

K. Daniel K., D. Hirsleifer and A. Subrahmanyam [7] show the diversified responses of individual investors on received information, as the

reason for the disclosure of the market paradoxes. Assumed lack of the strong effectiveness of the financial market is one of characteristics of this theory.

H. Hong and J. Stein [13] describe investment activity, as the game amongst investors applying fundamental analysis and investors applying technical analysis. This interaction of two rational theories produces such market phenomena, which are paradoxes from a point of view of the economic theory. Behavioural aspects are enclosed here in the choice of cognitive strategy.

R. Dacey and P. Zielonka [6] present behavioural approach similar to neoclassical one. They propose to describe behavioural reasons of economic decisions by means of subjective utility functions.

The behavioural present value model suggested at this work is applicable also in case of the strongly effective financial market. Thus proposed model differs from one given by K. Daniel, D. Hirsleifer and A. Subrahmanyam [7], who assume lack of strong effectiveness.

The probability distribution of return rate determined by behavioural present value is identical with the empirical distribution. The proposed model differs in it from the prospect theory [25], where a subjective transformation of the observed probability distribution is applicable.

Carried here considerations do not require applying the utility function. The proposed model differs in it from the model R. Dacey and P. Zielonka [6].

The behavioural present value is at imprecision risk which is a common feature of the proposed model and the model proposed by N. Barberis, A. Shleifer and R. Visny [1].

In the BPV model behavioural reasons of economic decisions allow to describe financial market action, as game between investors with different behavioural characteristics. Thus BPV model can be used in the theory proposed by H. Hong and J. Stein [13]

In articles [17] [18] [19] author put hypothesis on the possibility of applying Hiroto's sets as a formal tool of behavioural finances. In this paper, put hypothesis has been positive verified by estimating the return rate by using behavioural present value.

In this paper we formulated a formal behavioural present value model only. The next step should be empirical studies designed to estimate the parameters of this model.